	\documentclass[12pt,preprint2]{aastex} 
	\shorttitle{LS~I~+65~010} 
	\shortauthors{Grundstrom et al.} 
\begin{document} 
 
%%\received{} 
%%\accepted{} 
 
\title{Joint H$\alpha$ and X-ray Observations of Massive X-ray Binaries. I. \\ 
The B-Supergiant System LS~I~+65~010 = 2S~0114+650}  
 
\author{E. D. Grundstrom, J. L. Blair, D. R. Gies\altaffilmark{1},  
 W. Huang\altaffilmark{1,2}, \\ M. V. McSwain\altaffilmark{1,3,4},   
 D. Raghavan, R. L. Riddle\altaffilmark{1,5}, J. P. Subasavage,  
 D. W. Wingert\altaffilmark{1}} 
\affil{Center for High Angular Resolution Astronomy and \\ 
 Department of Physics and Astronomy,\\ 
 Georgia State University, P. O. Box 4106, Atlanta, GA 30302-4106; \\ 
 erika@chara.gsu.edu, blair.john@comcast.net, gies@chara.gsu.edu,   
 wenjin@astro.caltech.edu, mcswain@astro.yale.edu, raghavan@chara.gsu.edu,  
 riddle@tmt.org, subasavage@chara.gsu.edu, wingert@chara.gsu.edu} 
 
\altaffiltext{1}{Visiting Astronomer, Kitt Peak National Observatory, 
National Optical Astronomy Observatory, operated by the Association 
of Universities for Research in Astronomy, Inc., under contract with 
the National Science Foundation.} 
\altaffiltext{2}{Current address: Department of Astronomy, 
 California Institute of Technology, MS 105-24, 
 Pasadena, CA 91125} 
\altaffiltext{3}{Current Address: Astronomy Department, 
Yale University, New Haven, CT 06520-8101} 
\altaffiltext{4}{NSF Astronomy and Astrophysics Postdoctoral Fellow} 
\altaffiltext{5}{Current Address: Thirty Meter Telescope, 
 2632 E. Washington Blvd., Pasadena, CA 91107}
 
\author{A. M. Levine, R. A. Remillard} 
\affil{MIT Kavli Institute for Astrophysics and Space Research, \\
Massachusetts Institute of Technology, Cambridge, MA 02139-4307; \\ 
aml@space.mit.edu, rr@space.mit.edu} 
 
\slugcomment{Submitted to ApJ} 
\paperid{65757} 

%%%%%%%%%%%%%%%%%%%%%%%%%%%%%%%%%%%%%%%%%%%%%%%%%%%%%%%%%%%%%% 
 
\begin{abstract} 
 
We report on a three year spectroscopic monitoring program  
of the H$\alpha$ emission in the massive X-ray binary  
LS~I~+65~010 = 2S~0114+650, which consists of a B-supergiant  
and a slowly rotating X-ray pulsar.  We present revised orbital  
elements that yield a period of $P=11.5983 \pm 0.0006$~d and 
confirm that the orbit has a non-zero eccentricity $e=0.18 \pm 0.05$.  
The H$\alpha$ emission profile is formed in  
the base of the wind of the B-supergiant primary, and we show  
how this spectral feature varies on timescales that are  
probably related to the rotational period of the B-supergiant.   
We also examine the X-ray fluxes from the {\it Rossi X-ray  
Timing Explorer} All-Sky Monitor instrument, and we show that 
the X-ray orbital light curve has a maximum at periastron 
and a minimum at the inferior conjunction of the B-supergiant.  
We also show that the wind emission strength and the high  
energy X-ray flux appear to vary in tandem on timescales of  
approximately a year.  
 
\end{abstract} 
 
 
%%%%%%%%%%%%%%%%%%%%%%%%%%%%%%%%%%%%%%%%%%%%%%%%%%%%%%%%%%%%%%% 
 
\setcounter{footnote}{5} 
 
\section{Introduction}                              % Section 1 
 
Massive X-ray binaries consist of a massive, luminous star 
with a neutron star or black hole companion with an orbital separation  
small enough to power an accretion-driven X-ray source.  The mass donor 
is often a rapidly rotating Be star, and the neutron star companion  
accretes gas from the slowly outflowing circumstellar disk of the Be star 
\citep{coe00}.  A second kind of massive X-ray binary contains an OB supergiant as the  
mass donor, and mass transfer occurs by a Roche lobe overflow stream  
or stellar wind capture \citep{kap98}.   In both cases the mass transfer  
rate can be time variable and lead to large variations in the X-ray flux.  
The gas in the immediate vicinity of the mass donor will generally  
be dense and hot enough to be a source of H$\alpha$ emission, and our 
goal in this paper is to investigate how the mass loss fluctuations  
at the source (as observed in H$\alpha$) are related to the  
X-ray variations, particularly as measured with the All-Sky Monitor (ASM)   
instrument aboard the {\it Rossi X-ray Timing Explorer (RXTE)} satellite 
\citep{lev96}.  We present similar investigations of these co-variations in  
companion papers on the black hole binary Cyg~X-1 = HDE~226868 \citep{gie03} 
and the Be X-ray binaries LS~I~+61~303 \citep{gr06a} and HDE~245770 = A~0535+26  
and X~Per \citep{gr06b}.  
 
Our subject here is the massive X-ray binary system 2S~0114+650 (4U~$0114+65$) 
whose optical counterpart is LS~I~+65~010 (V662~Cas; HIP~6081; ALS 6517). 
The characteristics of the mass donor star in this system were  
determined in a spectroscopic and photometric study by \citet{rei96}.  
They show that the star is a supergiant with a spectral classification  
of B1~Ia.  They also suggest that the H$\alpha$ emission present  
in the spectrum forms in the stellar wind of the supergiant and  
that the expected wind parameters and accretion rate are consistent  
with the observed X-ray properties.  The orbital elements were  
first determined by \citet*{cra85} who found that the orbit has a 
small eccentricity and a period of 11.6~d.  The orbital  
period is also found in the X-ray flux variations \citep*{far06,wen06} 
along with two other periodic signals: a super-orbital period of 30.7~d (possibly related 
to the precession of the disk surrounding the neutron star; \citealt{far06}) 
and the pulsar period of 2.7~hr \citep*{fin92,hal00,koe03,bon05}.  
This is the longest spin period of any known X-ray pulsar, and  
\citet{li99} argue that the pulsar was probably born with a huge  
magnetic field (as a magnetar) in order to have spun down sufficiently  
within the lifetime of the mass donor star.  
 
Here we outline the results of an H$\alpha$ monitoring program  
on LS~I~+65~010 that spanned the years 1998 -- 2000.   We describe  
the spectroscopy in \S2 and present revised orbital elements 
in \S3.   We discuss the properties and temporal variations  
of the H$\alpha$ emission in \S4, and in \S5 compare them with  
the X-ray variations observed by {\it RXTE}/ASM. 
 
%%%%%%%%%%%%%%%%%%%%%%%%%%%%%%%%%%%%%%%%%%%%%%%%%%%%%%%%%%%%%%% 
 
\section{Observations}                              % Section 2 
 
We obtained 100 optical spectra of LS~I~+65~010 with the Kitt Peak  
National Observatory 0.9~m coud\'{e} feed telescope during six observing  
runs between 1998 August and 2000 December.  The details of  
the spectrograph arrangement for each run are listed in Table~1.  
All the spectra cover the region surrounding H$\alpha$, but 
they were made using two different spectral dispersions.   
During the first, fifth, and sixth runs, we used grating B  
(632 grooves mm$^{-1}$ with a blaze wavelength of 6000 \AA ~in second order) 
and obtained a resolving power of $R=\lambda / \delta \lambda \approx 9500$. 
The spectra from the other three runs were made with  
the lower dispersion RC181 grating   
(316 grooves mm$^{-1}$ with a blaze wavelength of 7500 \AA ; made in first order  
with a GG495 filter to block higher orders), which yielded an average  
resolving power of $R=4100 - 4400$.  Exposure times were usually 30 minutes,  
and the spectra generally have a S/N $\approx 100$ pixel$^{-1}$ in the continuum. 

\placetable{tab1}      % Table 1 - Observation Journal  
 
The spectra were extracted and calibrated using standard routines  
in {\it IRAF}\footnote{IRAF is distributed by the National Optical Astronomy  
Observatory, which is operated by the Association of Universities  
for Research in Astronomy, Inc., under cooperative agreement  
with the National Science Foundation.}, and then each continuum  
rectified spectrum was transformed onto a uniform heliocentric  
wavelength grid for analysis.  The removal of atmospheric lines was done by
creating a library of spectra from each run of the rapidly rotating 
A-star $\zeta$~Aql, removing the broad stellar features from these, 
and then dividing each target spectrum by the modified atmospheric 
spectrum that most closely matched the target spectrum in a selected 
region dominated by atmospheric absorptions.
 
%%%%%%%%%%%%%%%%%%%%%%%%%%%%%%%%%%%%%%%%%%%%%%%%%%%%%%%%%%%%%%%%%%%%%%% 
 
\section{Orbital Radial Velocity Curve}                     % Section 3 
 
The spectra from different runs cover different portions  
of the red spectrum (Table 1), but all record both the H$\alpha$ emission  
line and the \ion{He}{1} $\lambda 6678$ absorption line.  Thus, we will  
concentrate on these common features in the analysis.  We measured  
radial velocities by cross-correlating each spectrum with a standard  
template that was formed by averaging all the spectra from the  
fifth run (2000 October).   The spectral range for the cross-correlation  
was generally limited to the immediate vicinity of the \ion{He}{1} $\lambda 6678$ 
line, but we also included the region surrounding \ion{He}{1} $\lambda 7065$ 
for the spectra from the final two runs (2000 October and December).  
The positions of these two features in the template spectrum  
were measured by parabolic fits to the lower portions of the profiles,  
and we added their mean radial velocity, $-52.8\pm0.4$ km~s$^{-1}$, to each  
relative velocity to place them on an absolute scale.  We note that there are  
known line-to-line velocity differences in the spectrum that  
are related to line formation at different depths in the expanding  
atmosphere \citep{cra85,koe03}, and velocities based on the strong \ion{He}{1}  
lines are probably systematically more negative than those from lines 
formed deeper in the atmosphere.  We also made cross-correlation  
measurements of the strong interstellar line at 6613 \AA ~that we used  
to monitor the radial velocity stability of our spectra.  Our results 
are summarized in Table~2 (which appears in full in the electronic  
version) that lists the mid-point heliocentric  
Julian date of observation, the orbital phase, radial velocity,  
observed minus calculated residual from the fit, the measured velocity   
difference of the 6613 \AA ~interstellar line from its position in the  
template spectrum, and the equivalent width of the H$\alpha$ emission (\S4).   
The typical stellar radial velocity errors are about $\pm 1.7$ km~s$^{-1}$  
for the high dispersion spectra and about twice that value for the lower 
dispersion spectra (based upon the differences of closely spaced pairs  
of observations).  
 
\placetable{tab2}      % Table 2 - Radial velocities 
 
We determined orbital elements from these velocities using the  
non-linear least-squares fitting program of \citet{mor74}.  
We initially determined elements for both the new data and  
the archival measurements from \citet{cra85}, and while most  
of the elements were in good agreement, we found the systemic  
velocity $\gamma$ for the new set was 2.5 km~s$^{-1}$ lower than  
that based on the data from \citet{cra85}.  This is not  
surprising given the existence of systematic line-to-line  
velocity differences and that \citet{cra85}  
measured a different set of lines in the blue.  Consequently,  
we subtracted this value from all the archival velocities,  
and then made a new fit of the merged velocities.  The revised 
elements are presented in Table~3, and the radial velocity  
curve is shown in Figure~1.  The revised orbital elements  
are in reasonable agreement with the original values from \citet{cra85}  
that are also listed in Table~3.  The main differences are  
an order of magnitude improvement in the accuracy of the 
orbital period and a modest decrease in the orbital semiamplitude 
$K$.  
 
\placetable{tab3}      % Table 3 - Orbital elements  
 
\begin{figure} 
\begin{center} 
{\includegraphics[angle=90,height=12cm]{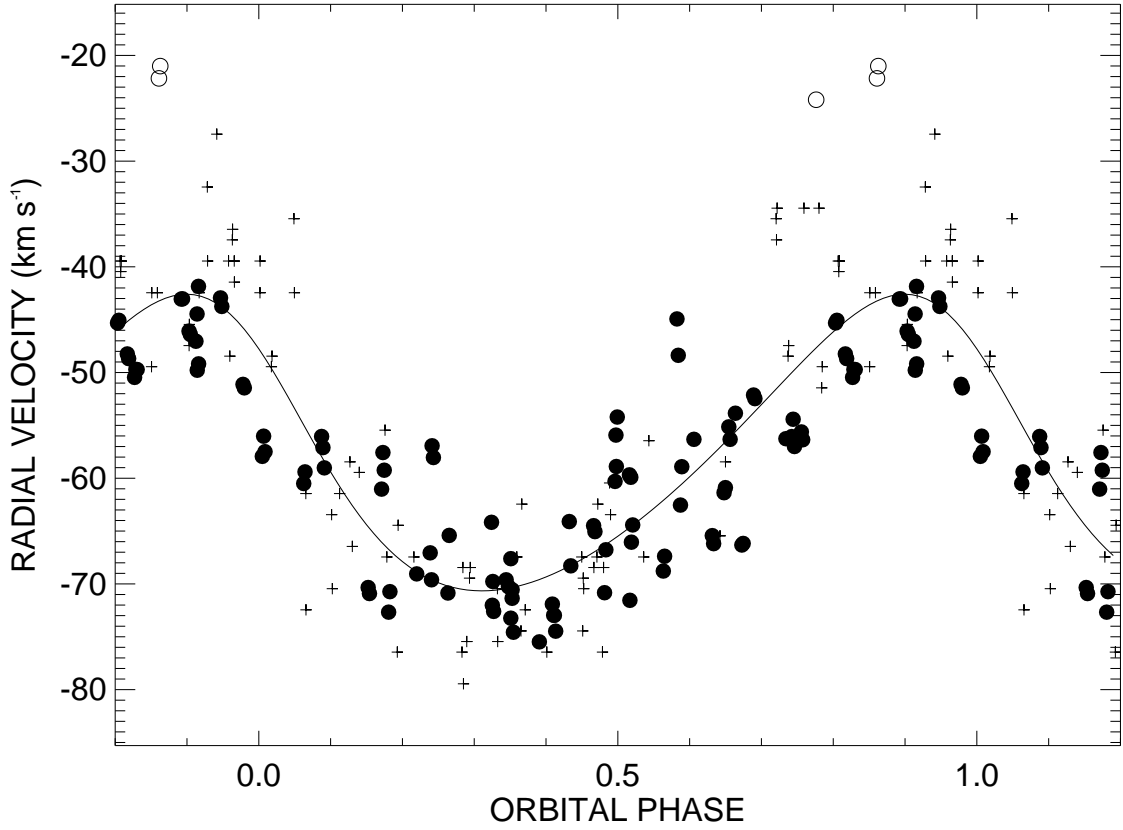}} 
\end{center} 
\caption{Calculated radial velocity curve ({\it solid line}) for LS~I~+65~010.   
The measured radial velocities are shown as filled circles. 
Open circles mark the measurements that were assigned zero weight (\S3), and  
plus signs indicate the measurements from \citet{cra85}.} 
\label{fig1} 
\end{figure}

\placefigure{fig1}     % Figure 1 - Radial velocity curve  
 
\citet{cra85} were unable to decide between circular and  
elliptical solutions, but it is now clear that a non-zero  
eccentricity is required.  We made fits assuming both circular  
and elliptical parameters, and the improvement in the residuals 
in the elliptical case would only occur by random chance with a  
probability of $0.8\%$ according to the test of \citet{luc71}. 
Thus, the eccentricity is significantly different from zero, and  
we report only the elliptical solution in Table~3.  The residuals  
from the fit (rms = 6.4 km~s$^{-1}$) are larger than those 
associated with the measurement errors (1.7 -- 3.4 km~s$^{-1}$), 
and, in fact, we zero-weighted three measurements 
that had exceptionally large deviations (Fig.~1).  
We suspect that this scatter is due to the microvariablility inherent  
to the atmospheres of early-type supergiants \citep{gie86}.

%%%%%%%%%%%%%%%%%%%%%%%%%%%%%%%%%%%%%%%%%%%%%%%%%%%%%%%%%%%%%%%%%%%%%%% 
 
\section{H$\alpha$ Variations}                              % Section 4 
 
The H$\alpha$ feature has always appeared in emission in past  
observations \citep{rei96,liu99,koe03} and emission was consistently   
present in our spectra as well.  However, the shape and strength of the  
profile varied on a timescale of days.  We show in Figure~2 
two example time sequences from our best sampled runs.    
A red-shifted emission peak is generally present and in about half  
of the spectra a blue-shifted absorption trough is also visible.  
There are indications in several cases of absorption features  
that appear to migrate bluewards to velocities of $\approx -400$  
km~s$^{-1}$.  These kinds of variations are common  
among B-supergiants \citep{ful97,mar00} and probably result  
from transient structures in the wind.  
Similar kinds of blueward absorption variations were also  
observed in the \ion{He}{1} $\lambda 5876$ line in the spectrum  
of LS~I~+65~010 by \citet{koe03}.  \citet{liu99} interpreted episodes of 
double-peaked H$\alpha$ emission as evidence of a second  
emission component from the vicinity of the neutron star companion,  
but our time sequences suggest that they are instead the result of  
varying stellar wind absorption.  
 
\begin{figure} 
\plottwo{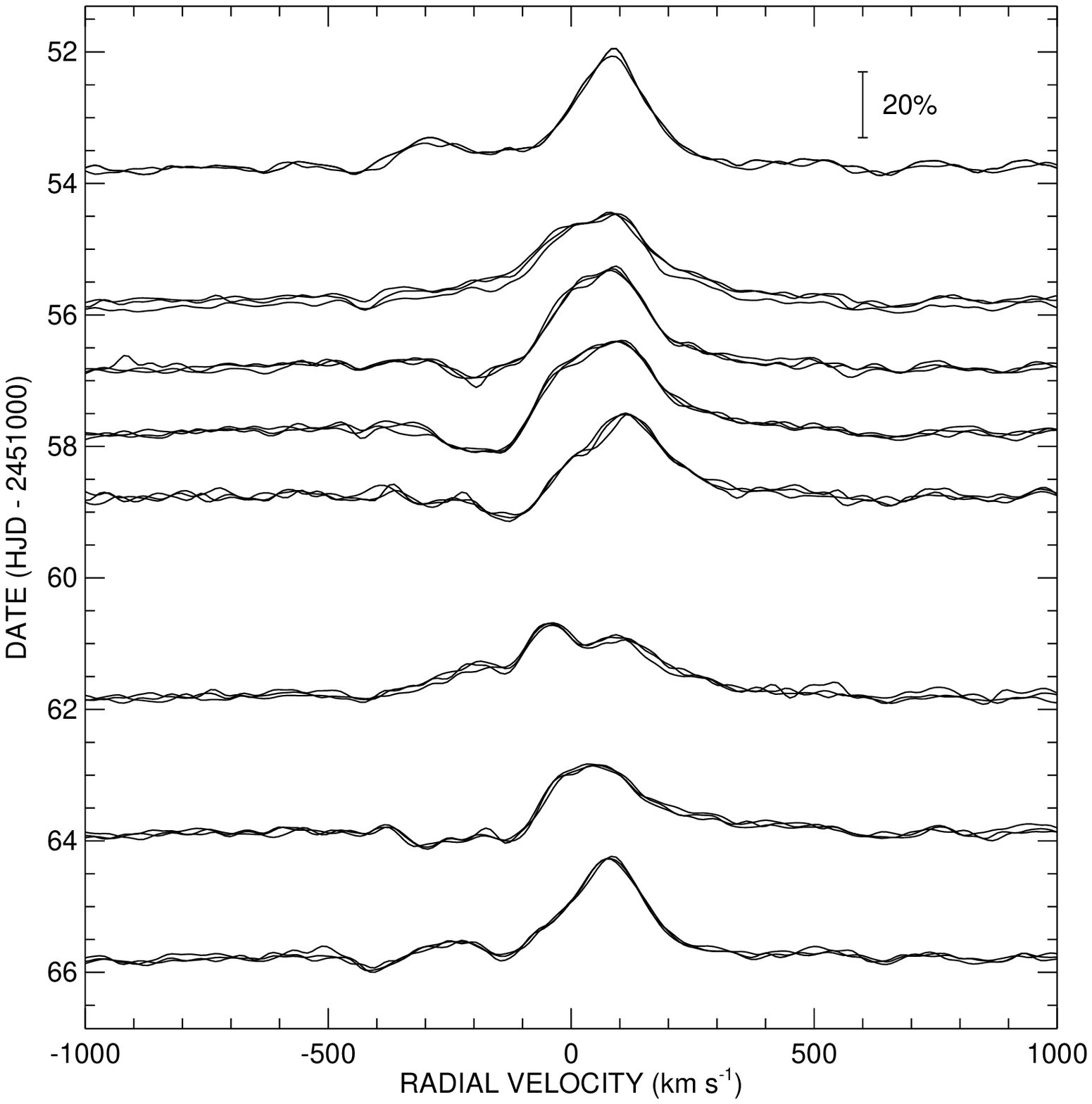}{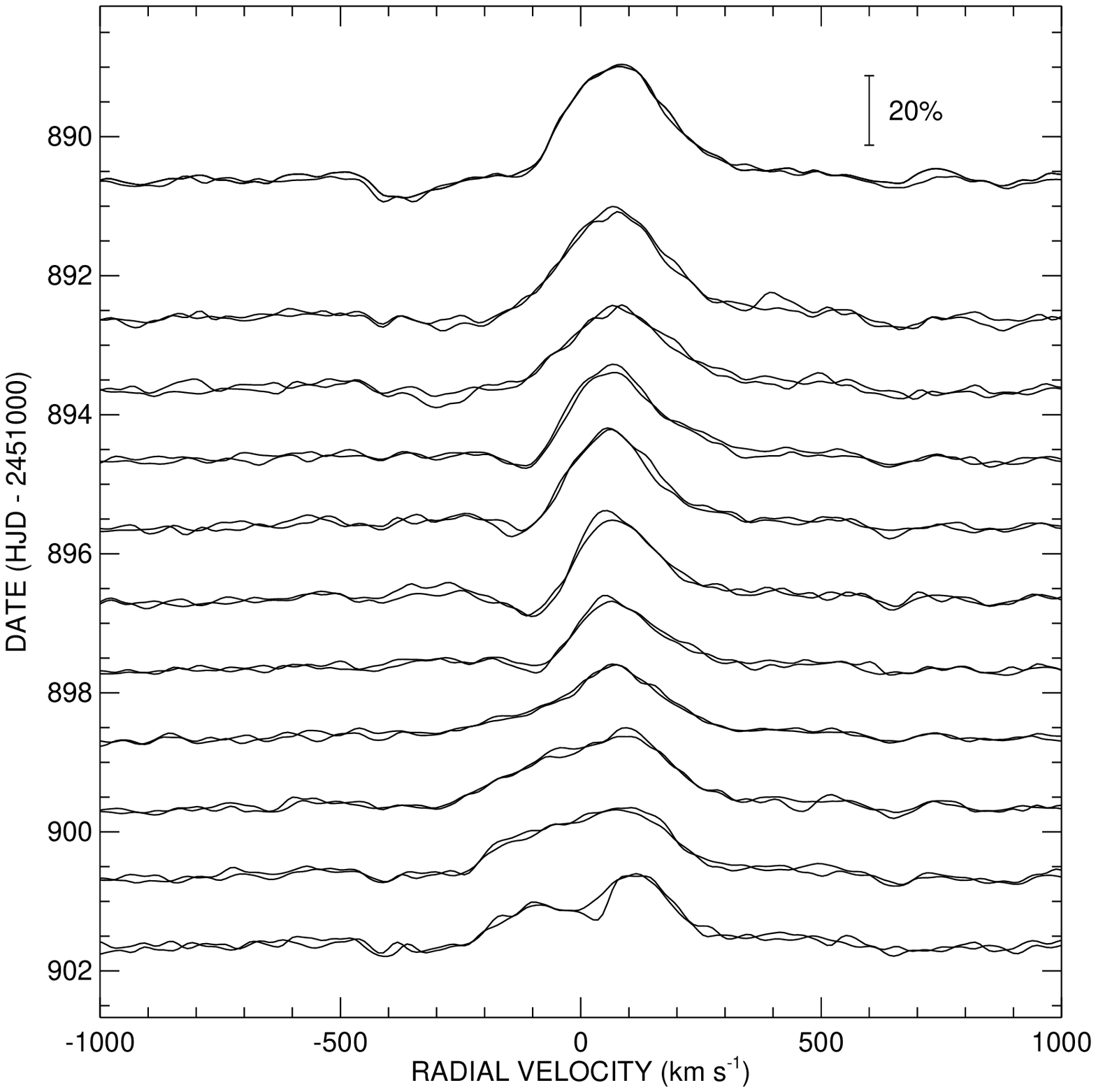} 
\caption{Temporal variations in the H$\alpha$ profile. 
The left panel shows the profiles from 1998 August 
while the right panel shows the variations in  
2000 December.  The vertical bar at upper right indicates 
the spectral intensity scale relative to a unit continuum. 
The spectra were smoothed with a Gaussian filter with  
FWHM = 30 km~s$^{-1}$.} 
\label{fig2} 
\end{figure}

\placefigure{fig2}     % Figure 2 - time sequence of profiles  
 
We measured emission equivalent widths for H$\alpha$ by making 
a simple numerical integration of line flux above the continuum,  
and our results are listed in the final column of Table~2.   
The typical measurement errors are approximately $\pm0.05$~\AA  
~for the higher dispersion spectra and about twice that for  
the lower dispersion spectra.  The resulting equivalents widths  
fall within the range reported in the past \citep{rei96,liu99}.   
We searched for any periodic signals in the time variations  
of the H$\alpha$ equivalent widths using the discrete Fourier  
transform and CLEAN algorithm\footnote{Written in the  
Interactive Data Language by A. W. Fullerton}  
\citep*{rob87}.  These results are summarized for each observing
season and for the entire data set in Figure~3.  
Most of the temporal variability appears to occur on timescales 
around 10~d (frequency of 0.1 cycles d$^{-1}$). 
We found no compelling evidence of periodic variability  
at either the orbital (11.6~d) or super-orbital period (30.7~d),  
and the $1\sigma$ upper limits for such cyclic variations 
are $<4\%$ and $<10\%$ for these two periods, respectively
(based upon sinusoidal fits at these fixed periods). 
The strongest signal in the power spectrum (among a number of 
candidate periods in the range between 6 and 12~d) occurs 
at a period of $9.72\pm 0.02$~d.  The false alarm probability 
that this peak is due to noise is small, $p_0 \approx N \exp (-z_0) = 0.0002$ 
where $z_0= 15.6$ is the ratio of the power peak to the variance and 
$N=1000$ is the number of quasi-independent frequencies tested over the 
range in physically interesting timescales between 0 and 1 cycle~d$^{-1}$ \citep{sca82}.  
Although H$\alpha$ variations occurred during each run on such timescales,  
there were large differences in the mean equivalent  
width between runs and in the detailed profile shape variations within runs 
that indicate the importance of other timescales as well.  
We suggest that this 10~d timescale represents a characteristic recurrence time 
for ephemeral structures to appear in a wind that is subject to longer term  
variations in mass loss rate.  If the structures are azimuthally  
distributed with respect to the star's spin axis, then this  
characteristic time may correspond to the stellar rotation period  
(see the case of the B-supergiant HD~64760 discussed by \citealt{ful97}
and the models for rotational modulation presented by \citealt{har00}).   
We can estimate the rotational period very approximately using the  
projected rotational velocity $V\sin i = 96\pm 20$ km~s$^{-1}$ and  
radius estimate $R/R_\odot = 37\pm 15$ \citep{rei96} plus the probable  
inclination range (see Fig.~6 below).  The timescale for the  
H$\alpha$ variations falls within the derived range of rotational period 
of 9 to 15~d.  
 
\begin{figure} 
\begin{center} 
\includegraphics[angle=90,height=12cm]{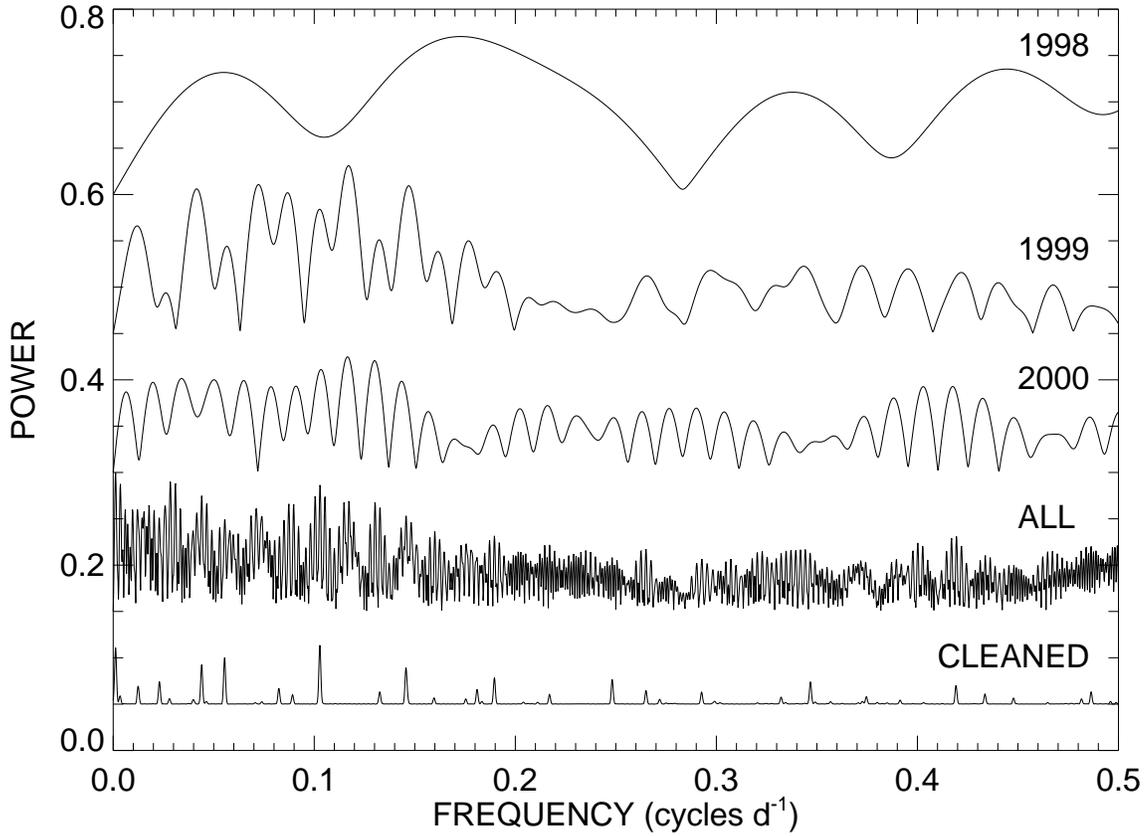} 
\end{center} 
\caption{Periodograms of the H$\alpha$ equivalent width 
variations for data from 1998, 1999, 2000, and all years 
({\it top to bottom}, with offsets of 
0.6, 0.45, 0.3, and 0.15, respectively). 
The lower plot shows the CLEANed periodogram (offset by 0.05) 
derived from a deconvolution of the full sample periodogram.} 
\label{fig3} 
\end{figure}

\placefigure{fig3}     % Figure 3 - H-alpha equivalent width periodograms 
 
%%%%%%%%%%%%%%%%%%%%%%%%%%%%%%%%%%%%%%%%%%%%%%%%%%%%%%%%%%%%%%%%%%%%%%% 
 
\section{X-ray Variations}                                  % Section 5 
 
The 2S~0114+650 system is one of some hundred targets that have been  
under continuous surveillance since 1996 March by the All-Sky Monitor  
on the {\it Rossi X-ray Timing Explorer} \citep{lev96}. 
This instrument records the X-ray flux across three energy bands  
in 90~s observation segments.  We extracted the flux measurements 
from 1996 March through 2006 May from the quick-look results provided  
by the {\it RXTE}/ASM team\footnote{http://xte.mit.edu/}.  
The source is relatively weak and there is a significant amount  
of scatter among the individual measurements.  However, since there  
are currently over 60,000 measurements available for 2S~0114+650,  
we can achieve a significant increase in S/N by averaging the  
fluxes into temporal or orbital-phase bins of interest.  
 
We first consider the long term variations in X-ray flux.  
We show in the upper panel of Figure~4 the high energy band fluxes  
for the duration of the {\it RXTE} mission binned into intervals  
equivalent to 12 binary orbits (139~d).  Each symbol shows the mean  
and standard deviation of the mean of the flux  
in the time bin.  These means and errors were taken from  
measurements made only with the Scanning Shadow Camera (SSC) 2  
and SSC3, since there is some indication of a long term trend  
of uncertain origin in the results from SSC1 for this particular 
source (possibly related to the calibration of gain changes in 
the proportional counter associated with SSC1).    
The lower panel illustrates the H$\alpha$ equivalent  
width measurements over the same dates (but with much more  
restricted time coverage).   There is a significant range in  
observed equivalent width during each run (\S4), but it appears that  
mean strength was largest in 1998 ($\sim$ HJD~2,451,060), 
smallest in 1999 ($\sim$ HJD~2,451,460), and intermediate 
in 2000 ($\sim$ HJD~2,451,860).  The binned X-ray fluxes in the top panel also 
show a similar kind of variation, which is suggestive that the time-averaged  
H$\alpha$ emission and X-fluxes appear to vary together.  
Taken at face value, this indicates that long term variations in  
the B-supergiant mass loss rate are reflected in the X-ray  
accretion flux.  A similar result was found for the massive  
X-ray binary and microquasar LS~5039 \citep{rei03,mcs04}.  
 
\begin{figure} 
\plotone{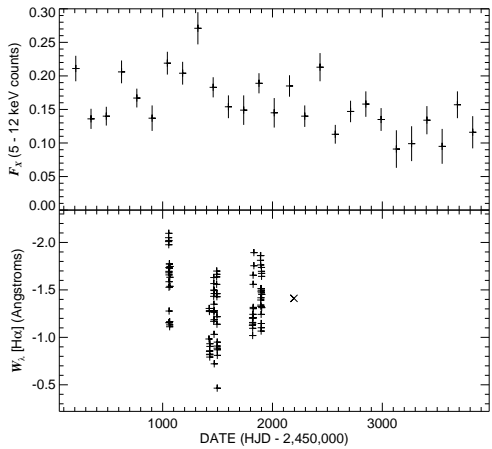} 
\caption{The long term variations in the {\it RXTE}/ASM high energy flux 
and H$\alpha$ equivalent width.  The top panel shows the ASM counts  
binned into intervals of 139~d (12 orbital periods) while the lower  
panel illustrates the H$\alpha$ emission strength.  The plus signs  
indicate our measurements while the single $\times$ sign is based on the  
observation from \citet{koe03}.} 
\label{fig4} 
\end{figure}

\placefigure{fig4}     % Figure 4 - X-ray and H-alpha long term variation 
 
Next we binned the ASM fluxes according to orbital phase from our  
spectroscopic solution (Table~3), and the light curves for  
the three energy bands are plotted in Figure~5.   
The orbital light curve is best defined in the 5 -- 12 keV band
where the observed counts are highest (top panel of Fig.~5; compare 
with the earlier light curve presented in Fig.~1 of \citealt{hal00}).   
We see that there is a clear maximum when the stars are closest  
near periastron (phase 0.0).  At that point in the orbit,  
the neutron star will encounter the densest and slowest  
portions of the supergiant's wind, and the wind accretion 
rate will attain a maximum \citep*{lam76}.   
The highest 5 -- 12 keV flux average actually  
occurs for the bin centered at phase 0.08, corresponding to  
a time some 22 hours after periastron.  This may indicate  
the interval between accretion and heating to X-ray emitting  
temperatures or it may represent the transit time for a wind  
density enhancement created by tidal effects at periastron  
to reach the vicinity of the neutron star.  
 
\begin{figure} 
\plotone{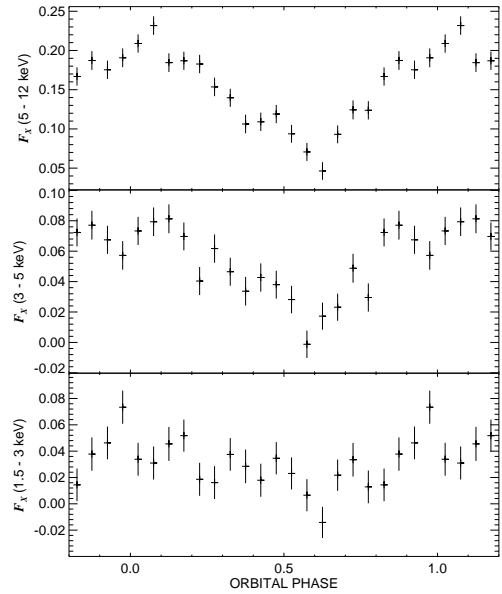} 
\caption{The {\it RXTE}/ASM fluxes binned according to orbital phase 
(phase 0.0 for periastron).  The panels show the bin mean and standard  
deviation of the mean of ASM counts for each of the three energy bands.} 
\label{fig5} 
\end{figure}

\placefigure{fig5}     % Figure 5 - X-ray orbital light curve  
 
The minimum in the X-ray light curve appears near phase 0.63,  
which is very close to the predicted phase of supergiant  
inferior conjunction at phase $0.65\pm 0.06$.   The X-ray flux  
does not vanish at this minimum, so we suspect that we are  
observing an atmospheric rather than a total eclipse \citep{hal00}.  
For example, \citet{wen99} show that in the case of Cyg~X-1, the  
decrease in X-ray flux observed when the supergiant is in the  
foreground is well explained by the attenuation of the X-ray flux by  
the wind of the supergiant.  We suspect that the same explanation  
applies to the case of the LS~I~+65~010 = 2S~0114+650 system  
since \citet{hal00} find evidence of an increase in column  
density near X-ray minimum.  
 
The kind of analysis made by \citet{wen99} of the wind attenuation  
as a function of orbital phase would also be profitable  
in this case and would presumably help place limits on the  
orbital inclination.  We show in Figure~6 the constraints on  
the masses of the two components in the mass plane diagram.  
The solid lines show the mass relations for given values of  
orbital inclination that are derived from the mass function  
($f(m)$ given in Table~3).  \citet{rei96} estimate that the supergiant 
mass and radius are $16\pm5 ~M_\odot$ and $37\pm15~R_\odot$,  
respectively, based upon their analysis of the optical and  
near-IR spectrum.  Note that this supergiant radius is much  
smaller than the star's Roche radius for the full range of 
acceptable inclination and mass ratio.   
The limiting line for the absence of  
full eclipses is shown by the dashed line in Figure~6  
(derived from the radius given above and the projected semimajor axis 
$a_1\sin i$ listed in Table~3).  Finally, we can place a weak constraint 
on a minimum inclination angle of $i=20^\circ$ by assuming that the  
supergiant rotates no faster than the critical rate (based on  
the given radius and observed projected rotational velocity 
$V\sin i = 96\pm20$ km~s$^{-1}$; \citealt{rei96}).   
This is a weak constraint because it is based  
upon the assumption that the supergiant's spin inclination is the  
same as the orbital inclination, but these inclinations could differ 
if the neutron star suffered an asymmetric kick at the time of the 
supernova explosion \citep{bra95}.  The shaded region
in Figure~6 shows the probable B-supergiant range together with
the observed range in neutron star mass \citep*{van95}, and 
a plus sign marks the nominal  
position for a Chandrasekhar mass neutron star ($1.4~M_\odot$).  
This best estimate solution has an inclination of $i=45^\circ$,  
just below eclipse limit near $i=50^\circ$, and this orientation  
is probably consistent with the atmospheric eclipse seen  
in the X-ray light curve.  A realistic model of the X-ray attenuation  
by the wind of the B-supergiant would provide a reliable estimate 
of the inclination and better constraints on the masses of both stars.  
 
\begin{figure} 
\begin{center} 
{\includegraphics[angle=90,height=12cm]{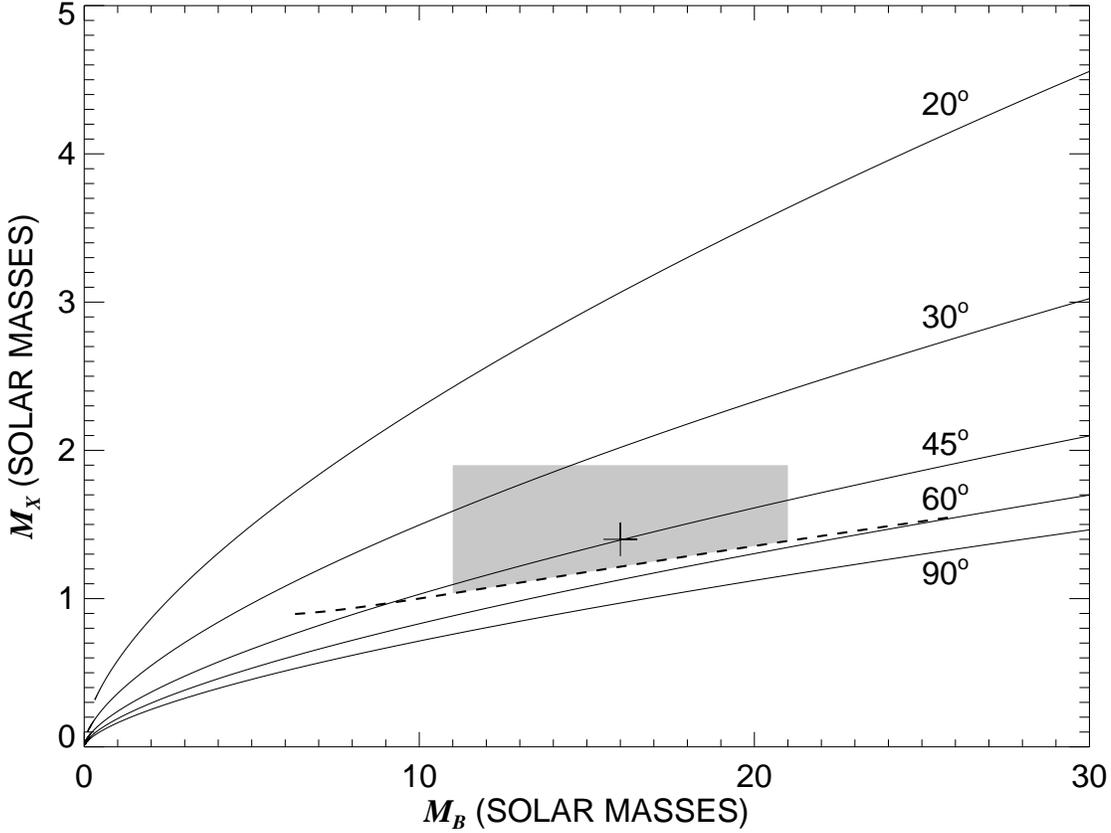}} 
\end{center} 
\caption{The relations between the supergiant mass $M_B$ and  
neutron star mass $M_X$ for LS~I~+65~010.  Lines of  
constant orbital inclination are constructed from the  
mass function and are labeled at the right hand side. 
The dashed line indicates the upper limit on inclination  
and the lower limit on X-ray source mass based upon the  
lack of eclipses.  The shaded region shows the  
probable solution space with the plus sign marking  
the favored solution.} 
\label{fig6} 
\end{figure}

\placefigure{fig6}     % Figure 6 - mass diagram 

%%%%%%%%%%%%%%%%%%%%%%%%%%%%%%%%%%%%%%%%%%%%%%%%%%%%%%%%%%%%%%%%%%%%%%% 
 
\acknowledgments 
 
We thank Daryl Willmarth and the staff of KPNO for their assistance  
in making these observations possible.  The X-ray results were  
provided by the {\it RXTE}/ASM teams at MIT and at the 
{\it RXTE} SOF and GOF at NASA's GSFC. 
This work was supported by the National Science Foundation 
under grants AST-0205297 and AST-0506573. 
Institutional support has been provided from the GSU College 
of Arts and Sciences and from the Research Program Enhancement 
fund of the Board of Regents of the University System of Georgia, 
administered through the GSU Office of the Vice President 
for Research.   
 
%%%%%%%%%%%%%%%%%%%%%%%%%%%%%%%%%%%%%%%%%%%%%%%%%%%%%%%%%%%%%%% 
 
% References 
 
%\clearpage 

%%%%%%%%%%%%%%%%%%%%%%%%%%%%%%%%%%%%%%%%%%%%%%%%%%%%%%%%%%%%%%% 
% Tables 
 
\clearpage 
 
% Table 1 - Journal of observations 
 
\begin{deluxetable}{cccc} 
\tablewidth{0pc} 
\tablecaption{Journal of Optical Spectroscopy \label{tab1}} 
\tablehead{ 
\colhead{Dates} & 
\colhead{Spectral Range} & 
\colhead{Resolving Power} & 
\colhead{Number} \\ 
\colhead{(HJD-2,400,000)} & 
\colhead{(\AA)} & 
\colhead{($\lambda/\triangle\lambda$)} & 
\colhead{of Spectra} } 
\startdata 
51053 -- 51065 & 6313 -- 6978       & 9530    &    23  \\ 
51421 -- 51429 & 5397 -- 6735       & 4050    &    11  \\ 
51464 -- 51469 & 5400 -- 6736       & 4100    &    14  \\ 
51491 -- 51497 & 5545 -- 6881       & 4400    &    14  \\ 
51817 -- 51830 & 6440 -- 7105       & 9500    &    16  \\ 
51890 -- 51901 & 6443 -- 7108       & 9500    &    22  \\ 
\enddata 
\end{deluxetable} 
 
\clearpage

% Table 2 - Velocity and equivalent width data 
 
\begin{deluxetable}{lccccc} 
\tabletypesize{\scriptsize} 
\tablewidth{0pt} 
\tablecaption{Radial Velocity and Equivalent Width Measurements\label{tab2}} 
\tablehead{ 
\colhead{Date} & 
\colhead{Orbital} & 
\colhead{$V_r$} & 
\colhead{$(O-C)$}	& 
\colhead{$\triangle V_{\rm ISM}$}	& 
\colhead{$W_\lambda ({\rm H}\alpha)$}	\\ 
\colhead{(HJD$-$2,400,000)} & 
\colhead{Phase} & 
\colhead{(km s$^{-1}$)} & 
\colhead{(km s$^{-1}$)}	& 
\colhead{(km s$^{-1}$)} & 
\colhead{(\AA )} } 
\startdata 
 51053.803\dotfill &  0.481 & $ -70.8$ & \phn   $-4.5$ & \phn   $-1.8$ &  $-$2.05 \\ 
 51053.824\dotfill &  0.483 & $ -66.8$ & \phn   $-0.5$ & \phn   $-0.4$ &  $-$2.01 \\ 
 51055.810\dotfill &  0.655 & $ -55.1$ & \phn\phs$1.1$ & \phn\phs$0.6$ &  $-$1.98 \\ 
 51055.832\dotfill &  0.656 & $ -56.3$ & \phn   $-0.2$ & \phn\phs$0.7$ &  $-$2.10 \\ 
 51055.917\dotfill &  0.664 & $ -53.9$ & \phn\phs$1.8$ & \phn\phs$3.9$ &  $-$2.02 \\ 
 51056.826\dotfill &  0.742 & $ -56.1$ & \phn   $-6.0$ & \phn\phs$2.4$ &  $-$1.70 \\ 
 51056.850\dotfill &  0.744 & $ -54.4$ & \phn   $-4.5$ & \phn\phs$2.3$ &  $-$1.68 \\ 
 51056.871\dotfill &  0.746 & $ -57.0$ & \phn   $-7.3$ & \phn\phs$2.2$ &  $-$1.74 \\ 
 51057.810\dotfill &  0.827 & $ -50.5$ & \phn   $-5.8$ & \phn\phs$1.0$ &  $-$1.53 \\ 
 51057.831\dotfill &  0.829 & $ -49.7$ & \phn   $-5.2$ & \phn\phs$2.0$ &  $-$1.59 \\ 
 51057.852\dotfill &  0.831 & $ -49.7$ & \phn   $-5.3$ & \phn\phs$2.1$ &  $-$1.66 \\ 
 51058.801\dotfill &  0.912 & $ -47.0$ & \phn   $-4.4$ & \phn\phs$3.6$ &  $-$1.15 \\ 
 51058.822\dotfill &  0.914 & $ -49.8$ & \phn   $-7.1$ & \phn\phs$2.8$ &  $-$1.16 \\ 
 51058.844\dotfill &  0.916 & $ -49.2$ & \phn   $-6.5$ & \phn\phs$2.4$ &  $-$1.28 \\ 
 51061.800\dotfill &  0.171 & $ -61.0$ & \phn\phs$5.0$ & \phn   $-2.0$ &  $-$1.77 \\ 
 51061.821\dotfill &  0.173 & $ -57.6$ & \phn\phs$8.6$ & \phn   $-0.1$ &  $-$1.77 \\ 
 51061.843\dotfill &  0.175 & $ -59.2$ & \phn\phs$7.1$ & \phn\phs$2.3$ &  $-$1.73 \\ 
 51063.887\dotfill &  0.351 & $ -73.2$ & \phn   $-2.9$ & \phn\phs$2.0$ &  $-$1.11 \\ 
 51063.908\dotfill &  0.353 & $ -71.4$ & \phn   $-1.0$ & \phn\phs$1.7$ &  $-$1.13 \\ 
 51063.930\dotfill &  0.355 & $ -74.6$ & \phn   $-4.3$ & \phn\phs$0.5$ &  $-$1.17 \\ 
 51065.811\dotfill &  0.517 & $ -71.5$ & \phn   $-6.9$ & \phn\phs$0.5$ &  $-$1.54 \\ 
 51065.835\dotfill &  0.519 & $ -66.0$ & \phn   $-1.5$ & \phn   $-0.5$ &  $-$1.63 \\ 
 51065.856\dotfill &  0.521 & $ -64.4$ & \phn\phs$0.0$ & \phn\phs$0.7$ &  $-$1.75 \\ 
 51421.915\dotfill &  0.220 & $ -69.0$ & \phn   $-0.2$ & \phn   $-3.2$ &  $-$0.98 \\ 
 51423.896\dotfill &  0.391 & $ -75.5$ & \phn   $-5.9$ & \phn   $-3.2$ &  $-$1.30 \\ 
 51425.898\dotfill &  0.563 & $ -68.8$ & \phn   $-6.7$ & \phn   $-6.6$ &  $-$1.27 \\ 
 51425.919\dotfill &  0.565 & $ -67.4$ & \phn   $-5.4$ & \phn   $-7.9$ &  $-$1.28 \\ 
 51426.879\dotfill &  0.648 & $ -61.4$ & \phn   $-4.7$ & \phn   $-0.4$ &  $-$0.82 \\ 
 51426.901\dotfill &  0.650 & $ -60.9$ & \phn   $-4.3$ & \phn   $-0.2$ &  $-$0.85 \\ 
 51427.879\dotfill &  0.734 & $ -56.3$ & \phn   $-5.7$ & \phn\phs$1.2$ &  $-$0.86 \\ 
 51428.838\dotfill &  0.817 & $ -48.2$ & \phn   $-3.1$ & \phn   $-1.8$ &  $-$0.79 \\ 
 51428.860\dotfill &  0.819 & $ -48.7$ & \phn   $-3.6$ & \phn   $-1.1$ &  $-$0.82 \\ 
 51429.836\dotfill &  0.903 & $ -46.1$ & \phn   $-3.5$ & \phn   $-1.5$ &  $-$0.93 \\ 
 51429.857\dotfill &  0.905 & $ -46.4$ & \phn   $-3.8$ & \phn\phs$1.0$ &  $-$0.90 \\ 
 51464.760\dotfill &  0.914 & $ -44.5$ & \phn   $-1.8$ & \phn   $-0.3$ &  $-$1.43 \\ 
 51464.783\dotfill &  0.916 & $ -41.9$ & \phn\phs$0.9$ & \phn\phs$1.0$ &  $-$1.17 \\ 
 51465.814\dotfill &  0.005 & $ -57.9$ & \phn   $-9.7$ & \phn   $-3.8$ &  $-$1.57 \\ 
 51465.835\dotfill &  0.007 & $ -56.0$ & \phn   $-7.6$ & \phn   $-5.6$ &  $-$1.63 \\ 
 51465.858\dotfill &  0.009 & $ -57.5$ & \phn   $-8.8$ & \phn   $-4.0$ &  $-$1.50 \\ 
 51466.772\dotfill &  0.087 & $ -56.0$ & \phn\phs$1.9$ & \phn   $-8.4$ &  $-$1.28 \\ 
 51466.794\dotfill &  0.089 & $ -57.1$ & \phn\phs$1.1$ & \phn   $-5.2$ &  $-$1.26 \\ 
 51466.816\dotfill &  0.091 & $ -59.0$ & \phn   $-0.6$ & \phn   $-3.4$ &  $-$1.35 \\ 
 51467.855\dotfill &  0.181 & $ -72.7$ & \phn   $-5.9$ & \phn   $-0.4$ &  $-$1.03 \\ 
 51467.876\dotfill &  0.183 & $ -70.7$ & \phn   $-3.9$ & \phn   $-2.2$ &  $-$1.19 \\ 
 51468.812\dotfill &  0.263 & $ -70.9$ & \phn   $-0.7$ & \phn\phs$1.3$ &  $-$1.46 \\ 
 51468.834\dotfill &  0.265 & $ -65.4$ & \phn\phs$4.8$ & \phn\phs$0.8$ &  $-$1.49 \\ 
 51469.829\dotfill &  0.351 & $ -67.6$ & \phn\phs$2.8$ & \phn   $-6.1$ &  $-$1.28 \\ 
 51469.850\dotfill &  0.353 & $ -70.5$ & \phn   $-0.2$ & \phn   $-2.1$ &  $-$0.72 \\ 
 51491.753\dotfill &  0.241 & $ -56.9$ &    \phs$12.7$ & \phn   $-2.9$ &  $-$1.56 \\ 
 51491.775\dotfill &  0.243 & $ -58.0$ &    \phs$11.7$ &    \nodata    &  $-$1.64 \\ 
 51492.713\dotfill &  0.324 & $ -64.2$ & \phn\phs$6.5$ & \phn\phs$3.0$ &  $-$1.70 \\ 
 51492.734\dotfill &  0.326 & $ -69.8$ & \phn\phs$0.8$ & \phn   $-0.8$ &  $-$1.66 \\ 
 51493.696\dotfill &  0.409 & $ -71.9$ & \phn   $-2.8$ & \phn\phs$3.0$ &  $-$1.46 \\ 
 51493.717\dotfill &  0.411 & $ -73.0$ & \phn   $-3.9$ & \phn\phs$3.4$ &  $-$1.43 \\ 
 51494.708\dotfill &  0.496 & $ -60.3$ & \phn\phs$5.4$ & \phn   $-2.3$ &  $-$1.14 \\ 
 51494.730\dotfill &  0.498 & $ -58.9$ & \phn\phs$6.7$ & \phn   $-4.5$ &  $-$1.22 \\ 
 51495.765\dotfill &  0.587 & $ -62.5$ & \phn   $-1.9$ & \phn\phs$1.1$ &  $-$0.91 \\ 
 51495.787\dotfill &  0.589 & $ -58.9$ & \phn\phs$1.6$ & \phn\phs$1.1$ &  $-$0.89 \\ 
 51496.757\dotfill &  0.673 & $ -66.3$ &       $-11.3$ & \phn   $-4.3$ &  $-$0.95 \\ 
 51496.778\dotfill &  0.675 & $ -66.2$ &       $-11.3$ & \phn   $-4.1$ &  $-$0.46 \\ 
 51497.718\dotfill &  0.756 & $ -55.6$ & \phn   $-6.5$ & \phn   $-2.5$ &  $-$0.87 \\ 
 51497.739\dotfill &  0.757 & $ -56.4$ & \phn   $-7.4$ & \phn   $-1.6$ &  $-$0.81 \\ 
 51817.702\dotfill &  0.344 & $ -69.6$ & \phn\phs$0.8$ & \phn   $-1.0$ &  $-$1.13 \\ 
 51817.739\dotfill &  0.348 & $ -70.3$ & \phn\phs$0.2$ & \phn   $-1.9$ &  $-$1.20 \\ 
 51818.722\dotfill &  0.432 & $ -64.1$ & \phn\phs$4.2$ & \phn   $-1.0$ &  $-$1.16 \\ 
 51818.746\dotfill &  0.434 & $ -68.3$ & \phn   $-0.0$ & \phn\phs$0.3$ &  $-$1.14 \\ 
 51819.693\dotfill &  0.516 & $ -59.7$ & \phn\phs$5.0$ & \phn\phs$1.0$ &  $-$1.10 \\ 
 51819.716\dotfill &  0.518 & $ -59.9$ & \phn\phs$4.7$ & \phn   $-2.4$ &  $-$1.02 \\ 
 51820.735\dotfill &  0.606 & $ -56.3$ & \phn\phs$3.2$ & \phn   $-0.0$ &  $-$1.21 \\ 
 51821.698\dotfill &  0.689 & $ -52.1$ & \phn\phs$1.7$ & \phn   $-0.6$ &  $-$1.20 \\ 
 51821.720\dotfill &  0.691 & $ -52.5$ & \phn\phs$1.2$ & \phn   $-1.5$ &  $-$1.24 \\ 
 51822.710\dotfill &  0.776 & $ -24.2$\tablenotemark{a} &    \phs$23.5$ & \phn\phs$1.3$ &  $-$1.21 \\ 
 51823.692\dotfill &  0.861 & $ -22.2$\tablenotemark{a} &    \phs$21.1$ & \phn\phs$0.2$ &  $-$1.65 \\ 
 51823.713\dotfill &  0.863 & $ -21.0$\tablenotemark{a} &    \phs$22.2$ & \phn\phs$0.7$ &  $-$1.56 \\ 
 51824.687\dotfill &  0.947 & $ -42.9$ & \phn\phs$0.8$ & \phn\phs$3.7$ &  $-$1.32 \\ 
 51824.708\dotfill &  0.949 & $ -43.8$ & \phn\phs$0.1$ & \phn\phs$4.1$ &  $-$1.30 \\ 
 51830.713\dotfill &  0.466 & $ -64.5$ & \phn\phs$2.5$ & \phn   $-1.2$ &  $-$1.89 \\ 
 51830.735\dotfill &  0.468 & $ -65.0$ & \phn\phs$1.9$ & \phn\phs$0.2$ &  $-$1.75 \\ 
 51890.622\dotfill &  0.632 & $ -65.4$ & \phn   $-7.6$ & \phn   $-2.0$ &  $-$1.81 \\ 
 51890.643\dotfill &  0.633 & $ -66.2$ & \phn   $-8.5$ & \phn\phs$0.8$ &  $-$1.86 \\ 
 51892.613\dotfill &  0.803 & $ -45.3$ & \phn\phs$0.6$ & \phn   $-2.9$ &  $-$1.86 \\ 
 51892.635\dotfill &  0.805 & $ -45.1$ & \phn\phs$0.8$ & \phn   $-1.4$ &  $-$1.76 \\ 
 51893.642\dotfill &  0.892 & $ -43.0$ & \phn   $-0.4$ & \phn   $-0.3$ &  $-$1.39 \\ 
 51893.664\dotfill &  0.894 & $ -43.0$ & \phn   $-0.4$ & \phn\phs$0.0$ &  $-$1.34 \\ 
 51894.640\dotfill &  0.978 & $ -51.1$ & \phn   $-5.4$ & \phn   $-2.0$ &  $-$1.42 \\ 
 51894.661\dotfill &  0.980 & $ -51.5$ & \phn   $-5.5$ & \phn   $-0.5$ &  $-$1.33 \\ 
 51895.620\dotfill &  0.063 & $ -60.5$ & \phn   $-5.5$ & \phn   $-0.7$ &  $-$1.51 \\ 
 51895.642\dotfill &  0.064 & $ -59.4$ & \phn   $-4.2$ & \phn   $-0.1$ &  $-$1.49 \\ 
 51896.662\dotfill &  0.152 & $ -70.3$ & \phn   $-5.7$ & \phn   $-0.5$ &  $-$1.07 \\ 
 51896.683\dotfill &  0.154 & $ -70.9$ & \phn   $-6.2$ & \phn\phs$0.5$ &  $-$1.24 \\ 
 51897.664\dotfill &  0.239 & $ -67.1$ & \phn\phs$2.5$ & \phn   $-1.5$ &  $-$1.15 \\ 
 51897.685\dotfill &  0.241 & $ -69.6$ & \phn\phs$0.0$ & \phn   $-0.6$ &  $-$1.10 \\ 
 51898.667\dotfill &  0.325 & $ -72.0$ & \phn   $-1.4$ & \phn   $-1.6$ &  $-$1.47 \\ 
 51898.688\dotfill &  0.327 & $ -72.6$ & \phn   $-2.0$ & \phn   $-1.4$ &  $-$1.50 \\ 
 51899.670\dotfill &  0.412 & $ -73.0$ & \phn   $-4.0$ & \phn\phs$0.3$ &  $-$1.70 \\ 
 51899.691\dotfill &  0.413 & $ -74.5$ & \phn   $-5.5$ & \phn   $-1.0$ &  $-$1.74 \\ 
 51900.664\dotfill &  0.497 & $ -55.9$ & \phn\phs$9.7$ & \phn\phs$1.0$ &  $-$1.67 \\ 
 51900.685\dotfill &  0.499 & $ -54.2$ &    \phs$11.3$ & \phn\phs$1.6$ &  $-$1.64 \\ 
 51901.649\dotfill &  0.582 & $ -44.9$ &    \phs$16.0$ & \phn\phs$1.2$ &  $-$1.31 \\ 
 51901.671\dotfill &  0.584 & $ -48.4$ &    \phs$12.5$ & \phn   $-1.1$ &  $-$1.45 \\ 
\enddata 
\tablenotetext{a}{Assigned zero weight in the orbital solution.} 
\end{deluxetable} 
 
\clearpage 
 
% Table 3 - Orbital elements 
 
\begin{deluxetable}{lcc} 
%\tabletypesize{\scriptsize} 
\tablewidth{0pc} 
\tablecaption{Orbital Elements\label{tab3}} 
\tablehead{ 
\colhead{Element}	&  
\colhead{\citet{cra85}}	& 
\colhead{This Work}	} 
\startdata 
$P$~(days)               \dotfill & $11.588 \pm 0.003$      & $11.5983 \pm 0.0006$  \\ 
$T$ (HJD--2,400,000)     \dotfill & $44134.9 \pm 0.7$       & $51825.3 \pm  0.5$ \\ 
$e$                      \dotfill & $0.16\pm0.07$           & $0.18 \pm 0.05$	\\ 
$\omega$ (deg)           \dotfill & $11 \pm 11$             & $51 \pm 17$  \\ 
$K$ (km s$^{-1}$)        \dotfill & $17 \pm 1$              & $14.0 \pm 0.7$  \\ 
$\gamma$ (km s$^{-1}$)   \dotfill & $-57 \pm 1$             & $-58.2 \pm 0.5$  \\ 
$f(m)$ ($M_\odot$)       \dotfill & $0.006 \pm 0.001$       & $0.0032 \pm 0.0005$  \\ 
$a_1 \sin i$ ($R_\odot$) \dotfill & $3.9 \pm 0.2$           & $3.16 \pm 0.16$  \\ 
r.m.s. (km s$^{-1}$)     \dotfill & $6.8$                   & $6.4$  \\ 
\enddata 
\end{deluxetable} 
 
%%%%%%%%%%%%%%%%%%%%%%%%%%%%%%%%%%%%%%%%%%%%%%%%%%%%%%%%%%%%%% 
 
% Figure captions 
 
\clearpage 
 
\input{epsf} 
% Figure 1 
 
\clearpage 
 
% Figure 2 
 
\clearpage 
 
% Figure 3 
 
\clearpage 
 
% Figure 4 
 
%\clearpage 
 
% Figure 5 
 
\clearpage 
 
% Figure 6 

%%%%%%%%%%%%%%%%%%%%%%%%%%%%%%%%%%%%%%%%%%%%%%%%%%%%%%%%%%%%%%% 
 
\end{document}